\pdfoutput=1

\documentclass{elsarticle}

\usepackage{graphicx}

\usepackage{color}
\usepackage{amsmath}
\usepackage{type1cm}
\usepackage{xspace} 

\usepackage{ulem}

\usepackage{euscript}

\newcommand{\D}[0]{\mathrm{d}}
\newcommand{\Li}[0]{\mathrm{Li}}

\newcommand{\abs}[1]{\lvert #1 \rvert}

\newcommand{\eps}[0]{\epsilon}

\journal{Physics Letters B}
\begin{document}
\begin{frontmatter}

\title{
\vspace{-4.0cm}
\begin{flushright}
		{\small {\bf IFJPAN-IV-2013-20}}   \\
		{\small {\bf SMU-HEP-13-26}}  \\
\end{flushright}
\vspace{0.5cm}
On regularizing the infrared singularities in QCD NLO splitting functions
with the new Principal Value prescription
} 

\author{ O.\ Gituliar$^a$, S.\ Jadach$^a$, A.~Kusina$^b$, M.\ Skrzypek$^a$ }
\address{$^a$ Institute of Nuclear Physics, Polish Academy of Sciences,\\
              ul.\ Radzikowskiego 152, 31-342 Cracow, Poland}
\address{$^b$ Southern Methodist University, Dallas, TX 75275, USA}


\begin{abstract}
We propose a modified use of the Principal Value prescription for
regularizing the infrared singularities in the light-cone axial gauge by
applying it to all singularities in the light-cone plus component of
integration momentum. The modification is motivated by and applied to the
re-calculation of the QCD NLO splitting functions for the purpose of Monte
Carlo implementations.
The final results agree with the standard PV prescription whereas contributions
from separate graphs get simplified.
\end{abstract}

\end{frontmatter}


\section{Introduction}
With the advent of the precision QCD measurements from the LHC there is an
interest in
the re-calculation of the QCD splitting functions at the NLO level, either
in order to construct a more precise, exclusive, Parton Shower Monte
Carlo (MC) algorithms or to improve the convergence of the logarithmic expansion
of PDFs \cite{Skrzypek:2011zw,Jadach:2012vs,Jadach:2011cr,deOliveira:2013maa}.
The physical interpretation of the evolution, necessary to construct the Parton
Shower MC, is best visible in the axial gauge in which the NLO calculations have
been done \cite{Curci:1980uw,Ellis:1996nn,Heinrich:1997kv,Bassetto:1998uv}. A
price to pay for the transparent physical picture is the appearance of the
spurious singularities associated with the axial denominator $1/(nl)$
where $n$ is the light-cone reference vector. These unphysical singularities
cancel at the end, but in the intermediate stages of the calculations one
has to regularize them somehow. The simplest way is to use the Principal Value
(PV) prescription \cite{Curci:1980uw,Ellis:1996nn,Hei98,Jadach:2011kc}. The
other option is the Mandelstam-Leibbrandt (ML)
prescription \cite{Mandelstam:1982cb, Leibbrandt:1983pj},
which is better justified from the field-theoretical point of view, but leads to
more complicated calculations, especially for the real-emission-graphs
\cite{Hei98}. Other methods of avoiding the problem of
spurious singularities can be found in \cite{Suzuki:1999cua,Bern:2004cz}.

The standard PV regularization is applied at the level of the Feynman rules to
the axial part of the gluon propagator:
\begin{align}
&  g^{\mu\nu} -\frac{l^\mu n^\nu + n^\mu l^\nu}{nl}
\to
g^{\mu\nu} -\frac{l^\mu n^\nu + n^\mu l^\nu}{[nl]_{PV}},
\;\;\;
\Bigl[\frac{1}{nl}\Bigr]_{PV} = \frac{nl}{(nl)^2 + \delta^2(pl)^2}
\label{PVprescr}
\end{align}
where $p$ is an external reference
momentum and $\delta$ is an infinitesimal regulator of the "spurious"
singularities.
Spurious singularities are artifacts of the gauge choice and are expected to
cancel completely
once the full set of graphs is added. On the other hand, apart from the axial
part of the propagators, there are also
other singularities in the $l_+=nl$ variable, associated with the Feynman part
of the propagator ($g^{\mu\nu}$) or phase space parametrization. In the
standard approach
\cite{Curci:1980uw,Ellis:1996nn,Heinrich:1997kv,Bassetto:1998uv} these
singularities are regularized by
means of dimensional regularization. As a consequence, in the final results for
single graphs we encounter both $\ln^2 \delta$ and $1/\epsilon^2$ terms.
This complicates calculations as well as makes results useless for the
stochastic simulations, which are supposed to be done in four dimensions.

In this note we propose a new way of using the PV
regularization. We show that the proposed scheme, called the NPV scheme, reproduces the
QCD NLO splitting functions correctly and in a simpler way. The
contributions from separate graphs are less singular in $\epsilon$, at the
expense of increased PV-regulated singularities%
\footnote{
   PV regularization is directly implementable in the MC computer codes.
         }, 
and the remaining
higher order singularities in most cases cancel separately in real and virtual
groups of diagrams.

\section{New PV prescription}
We propose to modify the PV prescription as follows: apply the PV
regularization of eq.\ (\ref{PVprescr}) to {\em all} the singularities in the
plus variable, not only to the axial denominators of the gluon propagators,
i.e.\
we propose to replace
\begin{align} 
d^ml\; l_+^{-1+\epsilon} \to d^ml\, \biggl[\frac{1}{l_+}\biggr]_{PV} 
\Bigl(1+\epsilon\ln l_+ +\epsilon^2 \frac{1}{2}\ln^2l_+ +\dots\Bigr),
\;\;\; l_+ = \frac{nl}{np}
\label{plusbar}
\end{align}
in the entire integrand and we keep the PV regulator $\delta$ small
but finite until the end of calculation.
The higher order $\epsilon$ terms are kept as needed.
In the
following we will denote this new scheme as the NPV prescription. 

The standard procedure of
introducing Feynman parameters, integrating out $m$-momentum and at last
integrating
out Feynman parameters, is not suitable for calculations in NPV scheme.
Instead, one has to isolate the integral
over the plus component of $m$-momentum and leave it for the very end.
The appropriate formulae are available in the literature \cite{Ellis:1996nn}
(see \cite{Hei98} for details of derivation). Let us quote here eq.\ (A.12) of
\cite{Ellis:1996nn} for the three-point integral with the kinematics
$p^2=(p-q)^2 = 0$, expanded to ${\cal O}(\epsilon^0)$ terms
\begin{align}
\label{EV}
\int & \frac{d^ml}{(2\pi)^m} 
         \frac{f(l_+)}{l^2 (l-q)^2(l-p)^2}
\notag
=  \\
= &\, \frac{-i}{16\pi^2q^2} \Bigl(\frac{4\pi}{-q^2}\Bigr)^{-\epsilon} 
\frac{\Gamma(1-\epsilon)}{-\epsilon}
  \biggl[
      \int_0^x dy f(l_+) z^{\epsilon}(1-z)^{\epsilon}
       \Bigl(1+2\epsilon\ln\frac{1-y}{1-z}\Bigr) \frac{1}{1-y}
\notag
\\ &
      +2\frac{\Gamma^2(1+\epsilon)}{\Gamma(1+2\epsilon)}(1-x)^{-\epsilon}
      \int_x^1 dy f(l_+) (1-y)^{-1+2\epsilon}
  \biggr],\;\;\; m=4+2\epsilon,
\end{align}
where
$x=q_+/p_+$, $ y=l_+/p_+$, $z=y/x=l_+/q_+$ and $f(l_+)$ is an arbitrary function
of the plus variable. The PV prescription enters through this
function, which can have end-point singularities at $y=0$ or $y=x$. There is
however also a singularity at $y=1$, in the last line of eq.\ (\ref{EV}), not
related to the axial function $f$. It is this singularity that is treated
differently: in the standard PV prescription it simply reads
$(1-y)^{-1+2\epsilon}$, whereas in our NPV one it is also regularized with
PV and becomes $(1-y)^{2\epsilon}[1/(1-y)]_{PV}$.
As a consequence even non-axial three point integrals are changed
and start to depend on the auxiliary vector $n$. Consider, for example, the
scalar integral 
\begin{equation} \label{eq:def-feyn}
  J_3^\mathrm{F}
  =
  \int \frac{\D^m l}{(2\pi)^m}
    \frac{1}{l^2 (q-l)^2 (p-l)^2}
\end{equation}
with the kinematical set-up: $p^2=(p-q)^2=0,\; q^2< 0$.
The PV regularization gives:
\begin{equation}
  J_3^\mathrm{F}
  =
  \frac{i}{(4\pi)^2 \abs{q^2}}
  \left(
    \frac{4\pi}{\abs{q^2}}
  \right)^{-\epsilon}
  \Gamma(1-\epsilon)
  \left(
    -\frac{1}{\eps^2} + \frac{\pi^2}{6}
  \right),
\label{eq:pv-f}
\end{equation}
whereas the new NPV prescription leads to:
\begin{equation}
\begin{split} \label{eq:our-f}
  J_3^\mathrm{F}
  = &
  \frac{i}{(4\pi)^2 \abs{q^2}}
  \left(
    \frac{4\pi}{\abs{q^2}}
  \right)^{-\epsilon}
  \Gamma(1-\epsilon)
  \bigg(
   - \frac{2 I_0 + \ln(1-x)}{\eps}
\\ &
    - 4 I_1 + 2 I_0 \ln(1-x) + \frac{\ln^2(1-x)}{2}
  \bigg),
\\
I_0 = & \int_0^1 dx\frac{1}{[x]_{PV}} = -\ln\delta +{\cal O}(\delta),
\\
I_1 = & \int_0^1 dx\frac{\ln x}{[x]_{PV}} = -\frac{1}{2} \ln^2\delta
-\frac{\pi^2}{24} +{\cal O}(\delta),
\end{split}
\end{equation}
where $x=q_+/p_+$ is the axial-vector-dependent parameter. 
As expected, eq.~\eqref{eq:our-f}
is free of double poles in $\eps$.
Instead, the $I_0/\epsilon$ and $I_1$ functions appeared.
The list of other three-point integrals needed for calculations of the NLO
splitting functions in the NPV scheme is given in the Appendix.

{\em Discussion:}
The use of the PV prescription has been criticized for a lack of a solid
field-theoretical basis, for example, for not preserving the causality
\cite{Nardelli:1989uj,Bassetto:1998uv}. On the other hand, it leads to
correct results for the NLO splitting functions
\cite{Curci:1980uw,Bassetto:1998uv}. The
singularities in $n$-direction are unphysical (spurious). As such, they must
cancel
at the end of the calculation once all the graphs are included. Therefore, as
argued in \cite{Curci:1980uw},  one
can employ a "phenomenological" PV recipe of how to deal with them in the
intermediate steps of the calculations. The proposed here new NPV scheme
follows the same justification: the
"non-spurious" IR singularities in plus-variable also cancel once the whole set
of graphs entering NLO splitting functions is added \cite{Ellis:1978sf}.
Therefore it is natural to extend the PV regularization and treat all the
singularities of the plus-variable on an equal footing.
Let us remark that separate regularization of the energy component
of the loop momentum is a known approach. For example in \cite{Heinrich:1999ak}
the singularities of the Coulomb gauge have been regularized by means of "split
dimensional regularization" in which the measure $d^m l$ is replaced by
$d^{2(\sigma+\omega)} l = d^{2\sigma} l_0 d^{2\omega}\vec{l}$.

On the technical level the NPV prescription simplifies the calculations -- one
does not need to
keep two types of regulators for the higher order poles. In the real emission
case the triple and double poles in $\epsilon$ vanish, replaced by $\ln\delta$,
the
calculations can be done in four dimensions
\cite{Jadach:2011kc} and there is no need of
cancelling these higher order poles between real and virtual graphs. 
The price to pay for these simplifications is that the non-axial integrals
become more complicated, as can be seen by comparing eqs.\
(\ref{eq:pv-f}) and (\ref{eq:our-f}).

\section{NLO splitting functions in the NPV scheme}
\begin{figure}[h]
  \centerline{
    \includegraphics[height=2.25cm]{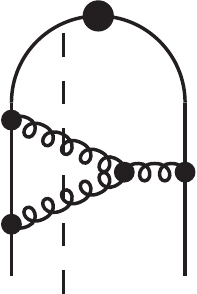}
    \hspace{1.5cm}
    \includegraphics[height=2.25cm]{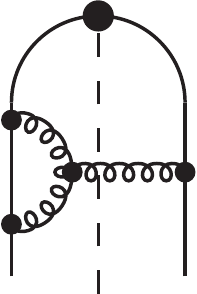}
    \hspace{1.5cm}
    \includegraphics[height=2.25cm]{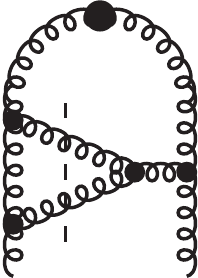}
    \hspace{1.5cm}
    \includegraphics[height=2.25cm]{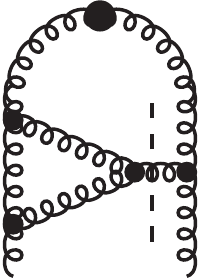}
  }
\centerline{          NS $(d_R)$ real\hskip 1.0cm NS $(d_V)$ virtual
          \hskip 1.1cm S $(d_R)$ real\hskip 1.3cm ~S $(d_V)$ virtual}
\caption{The real and virtual graphs of the type $"(d)"$, contributing to the
QCD NLO non-singlet (NS) and singlet (S) splitting functions.}
  \label{fig:graph-d}
\end{figure}

We are going to demonstrate now how the NPV scheme works in the calculations
of the
NLO quark-quark and gluon-gluon splitting functions and we show that
it reproduces the known final results of PV prescription
\cite{Curci:1980uw,Furmanski:1980cm}. 
More detailed results in the standard PV prescription can be found
in \cite{Hei98, Ellis:1996nn}. One can see there that in the standard PV
prescription the
triple poles, $1/\epsilon^3$, appear only in the four real and virtual
interference
graphs of the type $"(d)"$, shown in Fig.\ \ref{fig:graph-d}, both for the
non-singlet and singlet cases and only these graphs will be affected by the
change from PV to NPV prescription because the other, lower, $\epsilon$ poles
come from transverse- or minus-components of integration momenta.
We will discuss in turn these four contributions using the standard
formula relating the graphs $\Gamma$ and splitting functions $P$:
\begin{align}
\Gamma_{qq}(x,\epsilon)
 = \delta(1-x) 
  + \frac{1}{\epsilon}\biggl(
            \frac{\alpha_S}{2\pi} P_{qq}^{LO}(x)
       +\frac{1}{2}\Bigl( \frac{\alpha_S}{2\pi} \Bigr)^2  P_{qq}^{NLO}(x)
      +\dots \biggr)
  + {\cal O}(\epsilon^{-2}).
\notag
\end{align}
For more details we refer e.g.\ to \cite{Curci:1980uw,Hei98}.

\subsection{The virtual non-singlet graph $(d_V)$ in the NPV scheme}
In the NPV scheme the contribution from the non-singlet virtual graph $(d_V)$
to $\Gamma$ is
obtained using
the partial results of Ref.\ \cite{Hei98}:
the $T^{(d)V}$
function of eq.\ (3.152) and the counterterm $V^{(d)V}_{UV}$ of eq.\ (3.151).
The function  $T^{(d)V}$ is next evaluated with the help of the library of
integrals in the NPV scheme, given in the Appendix.
After subtracting $V^{(d)V}_{UV}$
and integrating over the one-particle real phase
space we end up with the result
\begin{align}
\begin{split}
\tilde \Gamma&_{qq}^{(d_V)}(x,\epsilon)
 =
\left(\frac{\alpha_S}{2\pi}\right)^2\Big(\frac{1}{2}
C_AC_F\Big)\biggl\{
     \frac{1}{\epsilon^2}P_{qq}(x) 
        \bigl(1 +\epsilon\ln(1-x)\bigr)\tilde{Z}_{d_V}
   - \frac{1}{\epsilon}\frac{1}{2}\frac{1+x}{1-x}
\\ &\,
   + \frac{1}{\epsilon}p_{qq}
     \biggl[ I_0  \bigl(
            2\ln x 
          + 2\ln(1-x)
          \bigr)
          - 6 I_1
          - \Li_2(1-x)
          + \ln^2x 
          - 3
          + \frac{8}{12} {\pi^2} 
      \biggr]
      \biggr\},
\label{td-2vd}
\end{split}
\end{align}
where the leading order kernel and renormalization constant are
\begin{align}
\tilde{Z}_{d_V}  = &\, 4I_0 +2\ln(1-x) +\ln x -\frac{3}{2},
\\
P_{qq}(x) =&\, p_{qq} +\epsilon(1-x),\;\;\;\; p_{qq}= \frac{1+x^2}{1-x}.
\end{align}

\subsection{The real non-singlet graph $(d_R)$ in the NPV scheme}
The non-singlet real contribution in the NPV scheme is given in eq.\ (3.48) of
ref.~\cite{Jadach:2011kc}.
Here, let us only compare its singular parts with the similar contributions in
the PV scheme. The singular terms, i.e.\ higher-order
pole terms and $I$-terms, are in the standard PV prescription 
(\cite{Hei98} Table 3.10):
\begin{multline} \label{d_hei}
  \frac{P_{qq}(x)}{\eps^2}
  -
  2 I_0 \frac{P_{qq}(x)}{\eps}
  +
  p_{qq}(x)
  \Bigl(
    - 2 I_1 + 4 I_0 + 2 I_0 \ln{x} - 2 I_0 \ln(1-x)
  \Bigr)
\end{multline}
and in the NPV prescription (\cite{Jadach:2011kc} eq.\ 3.48):
\begin{equation} \label{d_stj}
  p_{qq}(x)
  \Bigl(
    2 I_1 + 4 I_0 + 2 I_0 \ln{x} - 2 I_0 \ln(1-x)
  \Bigr).
\end{equation}
Those results are semi-inclusive, i.e.\ 
integration over one real momentum,  of the generic form 
$N(\epsilon)\int_0^{Q^2} d(-q^2) (-q^2)^{-1+2\epsilon}$
is not done.
Once performed, it will introduce additional $\eps$-pole.%
\footnote{
In eq.\ (3.48) in ref.~\cite{Jadach:2011kc}, instead of $1/(4\epsilon)$
pole, one finds $\ln (Q/q_0)$ where $q_0$ is
the lower limit of the integral 
$\int_{q_0}^{Q} d(|q|) (|q|)^{-1+4\epsilon}$, compare eq.\ (2.10) there.
         }
Here we clearly see the presence of the higher-order $\epsilon$ poles
in PV-formula and their
absence in NPV-formula, compensated by the change of the coefficient of the
$I_1$-term.

\subsection{Comments on non-singlet $d_V$ and $d_R$ result in the NPV scheme}
\label{sect:nscomp}
Let us compare the NPV results for the non-singlet graphs $(d)$
with the standard PV results available in the literature. Upon combining real
and virtual pieces, eq.\ (3.48) of ref.~\cite{Jadach:2011kc} and eq.\
(\ref{td-2vd}), we obtain in the NPV scheme
\begin{align}
\begin{split}
\tilde \Gamma&_{qq}^{(d_V)+(d_R)}(x,\epsilon)
 =
\left(\frac{\alpha_S}{2\pi}\right)^2\Big(\frac{1}{2}
C_AC_F\Big)\biggl\{
     \frac{1}{\epsilon^2}p_{qq}
        \tilde{Z}_{d_V}
\\ &\,
   + \frac{1}{\epsilon}p_{qq}
     \biggl[ I_0  \bigl(
            4\ln x 
          + 4\ln(1-x)
          \bigr)
          +  4I_0
          - 4 I_1
          + 3\ln x\ln(1-x)
\\ &~~~
     +\ln^2(1-x) +\frac{5}{2}\ln(1-x)
     + \frac{1}{2}\ln^2(x)
     - \frac{3}{4}\ln(x)
     - \frac{11}{2} 
     + \frac{1}{2} {\pi^2} 
\bigg]
\\&
     + \frac{1}{\epsilon} \biggl[\frac{1}{4}(1+x)\ln(x) - \frac{1}{2}(1-x)
     + (1-x) \biggl(4I_0 +2\ln(1-x) +\ln x \biggr)
\biggr] \bigg\}.
\end{split}
\label{eq:nsv+nsr}
\end{align}
This NPV result agrees with the results known from the
literature for the standard PV prescription. Namely, the $1/\epsilon^3$ and
$1/\epsilon^2$ $(d_V)$ and $(d_R)$ terms are given
in Table 3.10 of \cite{Hei98} and the $1/\epsilon$ terms are given in Table 1 of
\cite{Curci:1980uw} as a sum of real and virtual graphs. We observe, that:
\begin{itemize}
\item
The $1/\epsilon^3$ terms in NPV eq.\ (\ref{eq:nsv+nsr}) are {\em absent} both in
virtual and in real
graphs, as we expected. 
In the standard PV prescription these terms vanish only when real,
$(d)^{PV}_{R}$, and virtual, $(d)^{PV}_{V}$, contributions are added.
\item
The $1/\epsilon^2$ terms in NPV eq.\ (\ref{eq:nsv+nsr}) are of purely virtual
origin and are equal to the 
sum of the corresponding virtual $(d)^{PV}_{V}$ and real $(d)^{PV}_{R}$ terms
in the standard PV prescription.
This is a consequence of the absence of
$1/\epsilon^2$ terms in the real graph (d) in the NPV scheme, see eq.\
(\ref{d_stj}).
\item
The $1/\epsilon$ virtual plus real terms given in NPV eq.\ (\ref{eq:nsv+nsr})
agree with the known PV result. 
\end{itemize}
Let us note, that in the NPV
scheme there is no dependence on the upper integration limit 
$\ln Q^2$ and no "artifact"
terms $(\ln 4\pi -\gamma_E)$ are present, in none of the above $(d_V)$ and $(d_R)$
contributions. 
Of course, these "artifacts" would show up only in the plain MS scheme.
If $\overline{\text{MS}}$-like
scheme were used, these $(\ln 4\pi -\gamma_E)$ terms,
would be absent anyway. However, the dependence on $\ln Q^2$ would still be present
(if the PV prescription would have been used).

\subsection{The virtual singlet graph $(d_V)$ in the NPV scheme}
Next we turn to the graphs contributing to the singlet splitting
function, depicted in Fig.\ \ref{fig:graph-d}.
The calculation of the virtual graph proceeds as in the non-singlet
case. The 
counter term can be found e.g.\ in eq.\ (3.97) of \cite{Hei98}, the
corresponding $T^{(d)V}$ is not available in the literature.
The renormalized $\Gamma_{gg}^{(d_V)}$ is then calculated as the bare one
minus the counterterm integrated over the one particle phase space:
\begin{align}
\begin{split}
\label{svi}
\tilde \Gamma&_{gg}^{(d_V)} (x,\epsilon)
 =
           \left(\frac{\alpha_S}{2\pi}\right)^2C_A^2\frac{1}{2}\biggl[
           \frac{1}{\epsilon^2} P_{gg}
           \bigl(1+\epsilon\ln(1-x)\bigr)  \tilde{Z}_{GS}^V
\\ &
 + \frac{1}{\epsilon}P_{gg} \biggl(
            4 I_0 \ln(1-x)
          + 8 I_0 \ln(x)
          - 16 I_1
          + 4 \ln^2(x)
          + 12 \frac{\pi^2}{6}
          - \frac{134}{9} 
          \biggr)
       - \frac{1}{\epsilon}\frac{1}{3} x
   \biggr],
\end{split}
\end{align}
where
\begin{align}
\tilde{Z}_{GS}^V
= &
            12 I_0
          + 4 \ln(1-x)
          + 4 \ln(x)
          - \frac{22}{3} ,
\\
P_{gg} = & \frac{(1-x+x^2)^2}{x(1-x)}.
\end{align}

\subsection{The real singlet graph $(d_R)$ in the NPV scheme}
In order to complete the calculations we have computed the singlet real graph
$(d_R)$ in the NPV scheme. As in the non-singlet case, the calculation is less
complicated than in the standard PV scheme. Due to absence of higher order poles
it can be done in four dimensions, leading to
\begin{align}
\begin{split}
\label{sre}
\tilde \Gamma&_{gg}^{(d_R)}(x,\epsilon)
= -\left(\frac{\alpha_S}{2\pi}\right)^2C_A^2\frac{1}{2}
  \frac{1}{\epsilon}
  \biggl[
       P_{gg} \biggl( -4 I_1 
                        - 8 I_0 + 4 I_0 \ln(1-x) - 4 I_0 \ln(x)
\\ &
                         +2 \ln^2(1-x)
        +2 \ln^2(x) -4 \ln(x)\ln(1-x) -8 \ln(1-x) 
        +\frac{11}{3} \ln(x) +2 \frac{\pi^2}{6} +4
               \biggr)
\\ &
         +\ln(x) \Bigl(  
     \frac{11}{3} x^2 +\frac{23}{6}x +\frac{23}{6} +\frac{11}{3x}
                 \Bigr)
         -\frac{22}{3}x^2 + \frac{24}{3} x -\frac{25}{3} +\frac{22}{3x}
  \biggr].
\end{split}
\end{align}

\subsection{Comments on singlet $d_V$ and $d_R$ result in the NPV scheme}
\label{sect:scomp}
Now we compare
the singlet NPV results with the
corresponding
PV results from the literature.
Having added the real
and virtual graphs in the NPV prescription, eqs.\ (\ref{sre}) and
(\ref{svi}),
we obtain
\begin{align}
\begin{split}
\label{srv}
\tilde\Gamma&_{gg}^{(d_V)+(d_R)} (x,\epsilon)
 =
           \left(\frac{\alpha_S}{2\pi}\right)^2C_A^2\frac{1}{2}\biggl\{
           \frac{1}{\epsilon^2} P_{gg}
           \tilde{Z}_{GS}^V
\\ &
  + \frac{1}{\epsilon}\biggl[
  P_{gg} \biggl(
          12 I_0 \ln(x)
       +  12 I_0 \ln(1-x)
       +  8 I_0
       -  12 I_1
       +  2 \ln^2(1-x)
\\ &
       +  2 \ln^2(x)
       +  8 \ln(x) \ln(1-x)
       -  \frac{11}{3} \ln(x)
       +  \frac{2}{3} \ln(1-x)
       +  10 \frac{\pi^2}{6}
       -  \frac{170}{9}
         \biggr)
\\ &
       -  \Bigl( 
          \frac{11}{3} x^2 
       +  \frac{23}{6} x 
       +  \frac{23}{6} 
       +  \frac{11}{3x} 
          \Bigr) \ln(x)
       +\frac{22}{3} x^2
       -\frac{25}{3} x
       +\frac{25}{3}
       -\frac{22}{3x}
  \biggr]
  \biggr\}.
\end{split}
\end{align}
This NPV $(d_V)+(d_R)$ result agrees with the PV  
results from the literature:
in particular the $1/\epsilon^3$ and $1/\epsilon^2$ terms given
(separately for real and virtual graphs)
in Table 3.12 of ref.~\cite{Hei98} and the $1/\epsilon$ terms given
(only as a sum of real and virtual graphs)
in Table 4 column (cd) of ref.~\cite{Ellis:1996nn}.
The detailed comments to this comparison are identical as for the
non-singlet comparison of Sect.\ \ref{sect:nscomp}.

Since the other contributions to the NLO $P_{qq}$ and $P_{gg}$
splitting functions remain unchanged
while moving from the PV to NPV prescription, this completes the re-calculation 
of the NLO $P_{qq}$ and $P_{gg}$ splitting functions and demonstrates that the
final results are identical
in both schemes.

\section{Summary}
We proposed an extension of the use of the PV prescription in the light-cone
gauge, and we applied it to all the singularities in the plus component of the
integration momentum. We have shown that in the new NPV prescription the NLO
splitting functions, both non-singlet $P_{qq}$ and
singlet $P_{gg}$, are reproduced correctly.
The differences with respect to the PV scheme are present in partial results of
subset of four graphs, labelled $"(d)"$. The $1/\epsilon^3$ poles, present in
PV, are now replaced by $(1/\epsilon)\ln^2\delta$ etc. Therefore, the
calculations are easier, in particular the real graphs, now free of
$1/\epsilon^3$ and $1/\epsilon^2$ poles, can be calculated in four dimensions
and are usable for the Monte Carlo stochastic simulations, cf.\ eg.\
\cite{Jadach:2013dfd}.
The higher order poles cancel separately for real and for virtual components and
 neither real nor virtual contribution depend on the
scale $Q$ of the hard process. This is not true for the
standard
prescription -- in that case only the sum of real and virtual terms is
independent of $Q$. 
The drawback of the NPV prescription is that the non-axial integrals entering
calculations start to depend on the auxiliary vector $n$ and
become more complicated.

\section*{Acknowledgments}
This work is partly supported by 
 the Polish National Science Center grant DEC-2011/03/B/ST2/02632,
 the Polish National Science Centre grant UMO-2012/04/M/ST2/00240,
the U.S.\ Department of Energy
under grant DE-FG02-13ER41996 and the Lightner-Sams Foundation.
Two of the authors (S.J. and M.S.) are grateful for the
warm hospitality of the TH Unit of the CERN PH Division,
while completing this work.

\appendix
\section{Three-point integrals in NPV scheme}
We present the complete list of three-point integrals relevant for the
calculations of the NLO splitting functions which are modified in NPV with
respect to the standard PV prescription. The integrals are valid for the
specific kinematics:
$p^2=0,\;(p-q)^2=k^2=0,\; q^2<0$. The complete list of integrals in the PV
prescription can be found in Appendix A of \cite{Hei98}.%
\footnote{
\label{foot4}
The conventions used here are different from the ones of \cite{Hei98}:
$m=4 +2\epsilon$ (all poles in eqs.\ (\ref{first})--(\ref{last}) are of the IR
type).
The factor $(\mu_R^2)^{-\epsilon}/(2\pi)^m$ is
included in the normalization of integrals. The integration variable
$l$ is defined such that the denominators are $d^ml/(p-l)^2$, consequently,
the change of variable $l\to -l$ will result in additional $(-)$ sign for axial
denominator $1/nl$ and for each $l^{\mu_i}$ in the numerator, i.e.\ $J_3^A$
has different overall sign, but $J_3^{A\alpha}$ has the same
overall sign.
These changes of sign are compensated by the changes of sign in the
definitions of $J_3^\alpha$ and $J_3^A$
integrals in terms of form factors.

}
\begin{align}
\label{jgeneral}
\{J_3^A,\,J_{3}^{A\mu},\,J_{3}^{A\mu\nu}\}=
\int \frac{d^m l}{(2\pi)^m} \frac{\{1,\, l^\mu,\, l^\mu l^\nu\}}
{l^2 (p-l)^2 (q-l)^2}
\frac{1}{nl}.
\end{align}
The Feynman integrals, $J_3$, are similar but without the $1/(nl)$ term and
$J_3^{n3} = J_{3}^{\mu\nu\rho} n_\mu n_\nu n_\rho$ and so on.
The normalization $Q_\epsilon$, common to all the integrals, is defined as 
\begin{align}
Q_\epsilon(r)= i \frac{1}{(4\pi)^{2+\epsilon}}\Gamma(1-\epsilon)
\frac{(-r^2)^\epsilon}{(\mu_R^2)^\epsilon}.
\end{align}
Feynman integrals are defined as:
\begin{align}
J_3^{\alpha\beta}(q,p) =& \frac{Q_\epsilon(q)}{q^2} \Bigl(
   R_3 p^\alpha p^\beta 
  +R_4 q^\alpha q^\beta 
  +R_5 (q^\alpha p^\beta +p^\alpha q^\beta)
  +R_6 q^2 g^{\alpha\beta}         \Bigr),
\\
J_3^{\alpha}(q,p) = &\frac{Q_\epsilon(q)}{q^2} \Bigl(
   R_1 p^\alpha
  +R_2 q^\alpha           \Bigr),
\\
J_3^{n3}(q,p) = & -\frac{Q_\epsilon(q)}{q^2} (pn)^3 ( N_3 -R_0 ),
\\
J_3^{n2}(q,p) = & -\frac{Q_\epsilon(q)}{q^2} (pn)^2 ( N_2 -R_0 ),
\\
J_3^{n1}(q,p) = & -\frac{Q_\epsilon(q)}{q^2} (pn)^1 ( N_1 -R_0 ),
\\
J_3(q,p) =  & \frac{Q_\epsilon(q)}{q^2} R_0 .
\end{align}
Axial integrals are defined as:
\begin{align}
J_3^{A\alpha}(q,p) = & \frac{Q_\epsilon(q)}{q^2 pn} \Bigl(
   S_1 p^\alpha
  +S_2 q^\alpha
  +\frac{1}{2q^2 qn} S_3 n^\alpha           \Bigr),
\\
J_3^A(q,p) = & \frac{Q_\epsilon(q)}{q^2 pn} S_0 ,
\\
J_3^A(-q,k) = &\frac{Q_\epsilon(q)}{q^2 kn} U_0 .
\end{align}
Functions $N$ and the modified in
the NPV scheme functions ${R^{NPV}},\,{S^{NPV}}$ and
${U^{NPV}}$ read:%
\begin{align}
\label{first}
R^{NPV}_3 = &R^{NPV}_1 - \frac{1}{\epsilon} +3,  
\\
R^{NPV}_1 = &R^{NPV}_0 - \frac{2}{\epsilon} +4, 
\\
R^{NPV}_0 = &-\Bigl(-2 \frac{I_0}{\epsilon} -\frac{1}{\epsilon}
\ln(1-x) 
-4 I_1 +2 I_0 \ln(1-x) +\frac{\ln^2(1-x)}{2}\Bigr),
\\
\notag
U^{NPV}_0 = &-\Bigl(-\frac{3 I_0}{\epsilon} -\frac{3}{\epsilon}
\ln(1-x)
+\frac{1}{\epsilon} \ln(x) -5 I_1 +2 I_0 \ln(1-x) +I_0 \ln(x)
\\ &
             -\frac{1}{2}\ln^2(1-x)
          -2 \Li_2(1-x) +\frac{1}{2}\ln^2(x) +5 \frac{\pi^2}{6}
\Bigr), 
\\
\notag
S^{NPV}_0 = & \frac{3 I_0}{\epsilon} +\frac{1}{\epsilon} \ln(1-x)
-\frac{1}{\epsilon} \ln(x) +5 I_1 -2 I_0 \ln(1-x)
           -I_0 \ln(x) 
\\ &
-\frac{1}{2} \ln^2(x) -\frac{1}{2} \ln^2(1-x) -2 \Li_2(1-x)
-\frac{\pi^2}{6},
\\
\notag
S^{NPV}_1 = & \frac{2 I_0}{\epsilon} +\frac{1}{\epsilon} \ln(1-x) +4
I_1 -2
I_0 \ln(1-x) -\frac{1}{2} \ln^2(1-x)
\\ &
  +\frac{1}{\epsilon}\frac{x}{(1-x)} \ln(x) +\frac{x}{(1-x)} \Li_2(1-x),
\\
N_3 = &- \frac{1}{\epsilon}\Bigl( \frac{x^3}{3} + \frac{x^2}{2}
+{x} - \frac{11}{3} \Bigr)
        + \Bigr(  \frac{13}{18} x^3 +\frac{4}{3} x^2 
                 +\frac{11}{3} x -\frac{85}{9} \Bigr),
\\
N_2 = &- \frac{1}{\epsilon}\Bigr( \frac{x^2}{2} +{x - 3}
\Bigr)
        + ( x^2 +3 x -7 ),  
\\
N_1 = &- \frac{ x - 2 }{\epsilon}
        + 2 ( x -2 ).
\label{last}
\end{align}
The remaining $R,\, S$ and $U$ functions are identical in the PV and NPV schemes
and can be found in \cite{Hei98}, see also footnote \ref{foot4}.

\bibliographystyle{elsarticle-num}
\bibliography{radcor}

\end{document}